\documentclass[preparint]{elsarticle}

\usepackage{longtable}
\usepackage{lineno,hyperref}
\usepackage{amsmath}
\usepackage{float}
\usepackage{graphicx}
\usepackage{subfigure}
\usepackage{caption}
\modulolinenumbers[5]

\journal{Journal of \LaTeX\ Templates}

\begin{document}

\begin{frontmatter}

\title{LFGCF: Light Folksonomy Graph Collaborative Filtering for Tag-Aware Recommendation}
\author{Yin Zhang, XianJun Wu, Yan Zhang, LiGang Dong}
\address{School of Information and Electronic Engineering(Sussex Artificial Intelligence Institute), Zhejiang Gongshang Univeristy, Hangzhou, China}

\author{Can Xu, Weigang Wang}
\address{School of Statistics and Mathematics, Zhejiang Gongshang Univeristy, Hangzhou, China}

\begin{abstract}
Tag-aware recommendation is a task of predicting a personalized list of items for a user by their tagging behaviors. It is crucial for many applications with tagging capabilities like last.fm or movielens. Recently, many efforts have been devoted to improving Tag-aware recommendation systems (TRS) with Graph Convolutional Networks (GCN), which has become new state-of-the-art for the general recommendation. However, some solutions are directly inherited from GCN without justifications, which is difficult to alleviate the sparsity, ambiguity, and redundancy issues introduced by tags, thus adding to difficulties of training and degrading recommendation performance.

In this work, we aim to simplify the design of GCN to make it more concise for TRS. We propose a novel tag-aware recommendation model named Light Folksonomy Graph Collaborative Filtering (LFGCF), which only includes the essential GCN components. Specifically, LFGCF first constructs Folksonomy Graphs from the records of user assigning tags and item getting tagged. Then we leverage the simple design of aggregation to learn the high-order representations on Folksonomy Graphs and use the weighted sum of the embeddings learned at several layers for information updating. We share tags embeddings to bridge the information gap between users and items. Besides, a regularization function named TransRT is proposed to better depict user preferences and item features. Extensive hyperparameters experiments and ablation studies on three real-world datasets show that LFGCF uses fewer parameters and significantly outperforms most baselines for the tag-aware top-N recommendations.
\end{abstract}

\begin{keyword}
Recommendation systems \sep  Tag-aware collaborative filtering \sep Graph neural networks \sep Knowledge graph
\MSC[2022] 05-23\sep  editon: 1.2
\end{keyword}

\end{frontmatter}

\section{Introduction}
\label{sec:introducion}
Tag-aware recommendation systems (TRS), to better depict user preferences and item features, have been widely deployed to bridge the information gap between users and items \cite{hutchison_information_2006, bischoff_can_2008}. The fundamental factor of TRS is to provide a folksonomy, where users can freely assign tags to items that they interacted with (e.g., movies, music, and bookmarks) \cite{rendle_learning_2009}. These tags are composed of concise and comprehensive words or phrases that reflect users' subjective preferences and items' characteristics \cite{chen_tgcn_2020}. From this perspective, tags in folksonomy can bridge collaborative information between users and items. By exploring these collabortaive information in the tagging procedure, TRS is able to provide personalized item lists for users \cite{zuo_tag-aware_2016,li_tag-aware_2019,chen_tgcn_2020}. Therefore, folksonomy records can be introduced into recommendation systems to enhance interpretability and improve recommendation quality.

A common paradigm is to transform tags into a generic feature vector and feed them into feature-based models to integrate auxiliary folksonomy records. For example, CFA\cite{zuo_tag-aware_2016} used the sparse autoencoder (SAE) to obtain tag-based user latent representations and combines those with user-based collaborative filtering (CF). Besides, DSPR\cite{xu_dspr_2016} and HDLPR\cite{xu_hdlpr_2017} leveraged the multi-layer perceptron (MLP) to process such sparse feature vectors and extract abstract user and item representations, AIRec\cite{chen_airec_2021} provided a hybrid user model with hierarchical attention networks, which can depict user implicit preferences from explicit folksonomy records. Some researchers have organized the folksonomy records as a graph in recent years, utilizing graph neural networks (GNN) for TRS. TGCN\cite{chen_tgcn_2020} used graph convolutional networks (GCN) for TRS and outperformed other state-of-the-art TRS models. TA-GNN\cite{huang_tagnn_2021} leveraged two graph attention networks for embeddings aggregation and achieve higher quality recommendations as well. 

Although the above method improves the recommendation performance in TRS, it comes with some unacceptable issues resulting from the sparsity of data and the redundancy and ambiguity of tags \cite{shepitsen_personalized_2008}. To be specific, the \textbf{sparsity} issues from most users assigning a few amounts of tags to the items they interacted with. The \textbf{redundancy} and \textbf{ambiguity} issues from the fact that several tags have the same or different meaning due to the lack of contextual semantics. For example, they are owing to the diversity in writing or expression styles of users. Some tags with different forms have the same meanings and indicate similar preferences, such as ``world war 2" and ``ww2" being two different tags in folksonomy, but usually assigned to the same movie ``Schindler's List" by users. Moreover, some tags consist of polysemous words, which means they have different understanding in different contexts. The tag ``apple" could be misunderstood as a technology company by most tech-enthusiasts rather than a kind of fruit. Those issues increase training difficulty and finally degrade the recommendation systems' effectiveness. 

Several recent works which introduced GNN to TRS have demonstrated its effectiveness to deepen the use of subgraph with high-hop neighbors, such as TGCN\cite{chen_tgcn_2020}, GNN-PTR\cite{chen_graph_2020} and TA-GNN\cite{huang_tagnn_2021}. However, those works have shown promising results; we argue that its designs are rather burdensome, directly inherited from GCN without justification.

This paper focuses on the issues mentioned above and proposes a GCN-based recommendation model for TRS. Specifically, we construct two sets of edges among user tagging records. The first set of edges reflects the initiative interaction between users and tags, and the second set of edges reflects passive interaction between items and tags. Therefore, the Folksonomy Graph (FG) can be constructed based on the above edges set. It is worth mentioning that relationships between users and items are removed because part of tagging behavior is a negative interaction. For example, some users assign the tag ``worst movie ever" for some items. In this case, the FG is constructed based on the tagging and tagged information. Fig.\ref{folksonomy} illustrates an example. Then, we propose a Light Folksonomy Graph Collaborative Filtering  (LFGCF) inspired by LightGCN\cite{he_lightgcn_2020}, on which the Graph Convolutional Networks integrate topological structure for representation learning and share tag representations, which bridge the information gaps between users and items. Finally, we design a regularization function named TransRT to model the representation and perform joint learning with the recommendation task from a user tagging and item tagged perspective.

The main contributions of this paper are summarized as follows:

\begin{itemize}
    \item We construct FG based on the user tagging records and item tagged records, respectively, which reflect users' preferences and items' characteristics. The interaction between users and items is not used to construct the graph structure;
    \item We leverage the GCN for representation learning, which is specific light designed for TRS, and jointly optimize TransRT for the Top-K recommendation task;
    \item We perform extensive experiments on three public datasets, demonstrating improvements of LFGCF over state-of-the-art GCN methods for the tag-aware Top-K recommendation.
\end{itemize}

The rest of this paper is organized as follows. Section~\ref{sec:related} present the related words about TRS and GNN-based recommendation systems. Section~\ref{sec:material} gives some background, including problem formulation and definition FG. In Section~\ref{sec:method}, we propose a recommendation method based on GCN is described in detail. Section~\ref{sec:experiment} reports the hyperparameters experiments and ablation studies results. Finally, we conclude our contributions and give further work directions in Section~\ref{sec:conclusions}.

\section{Related Works}
\label{sec:related}
\subsection{Tag-aware recommendations}
Collaborative tagging recommendation allows users to assign tags freely on all kinds of items and then utilizes them to provide better-personalized recommendations. However, just like collaborative filtering recommendation, collaborative tagging recommendation suffers from the sparsity of interactions. Nevertheless, the redundancy and the ambiguity greatly compromise the performance of recommendations. By incorporating tagging information into collaborative filtering, Zhen et al. \cite{zhen_tagicofi_2009} came up with TagiCofi, which managed to characterize users' similarities better and achieved better recommendations. To tackle the redundancy of tags, Shepisten et al. \cite{shepitsen_personalized_2008} used hierarchical clustering on social tagging systems. Peng et al. \cite{peng_collaborative_2010} further integrated the relationship between users, items, and tags and then came up with a Joint Item-Tag recommendation. Zhang et al. \cite{zhang_personalized_2010, zhang_solving_2012} incorporated a user-tag-item tripartite graph, which turned out effective for improving the accuracy, diversity, and novelty of recommendations. FolkRank++ was brought up by Zhao et al. \cite{zhao_folkrank_2021} to dig out the similarities among users and items fully. Focusing on personalized ranking recommendations in a collaborative tagging system, Rendle et al. \cite{rendle_learning_2009} introduced matrix factorization and brought up RTF, which optimized ranking in personalized recommendations. Later, enlightened by BPR \cite{rendle_bpr_2009}, Li et al. \cite{li_tag-aware_2019} came up with BPR-T for tag-aware recommendations. With the development of deep learning, Zuo et al. \cite{zuo_tag-aware_2016} used deep neural networks to extract tag information, which helped alleviate the sparsity, redundancy, and ambiguity of tags and generate accurate user profiles.

Despite the fact that much effort has been devoted to tag-aware recommendations, the implicit pattern under user tagging behavior are not fully extracted. We introduce the modified knowledge graph algorithm to tackle this problem.

\subsection{GNN-based recommendations}
Recent years have witnessed rapid development in graph neural networks, which perform excellently in node classifications and edges predictions. Berg et al. \cite{berg_graph_2017} brought up GCMC, in which autoencoder was used in a user-item bipartite graph to generate expressive embeddings. Ying et al. \cite{ying_graph_2018} focused on web-scale recommendation systems and came up with a GCN-based algorithm: PinSage. Results showed that Pinsage had excellent robustness and could generate high-quality recommendations. Wang et al. \cite{wang_neural_2019} incorporated GCN and came up with NGCF. Benefited by the multi-hop neighborhood connectivities, NGCF achieved good recommendations performance. Later in the research of He et al. \cite{he_lightgcn_2020}, it is proven that some designs in NGCF are  burdensome and compromising recommendation performance. So LightGCN was brought up to simplify the model design, achieving better recommendations. Focusing on click-through rate (CTR) predictions, Li et al. \cite{li_fi-gnn_2019} came up with Fi-GNN, in which the gated recurrent units (GRU) was used for information aggregation. DG-ENN was brought up by Guo et al. \cite{guo_dual_2021} for the same task. By incorporating the attribute graph and the collaborative graph, DG-ENN was able to alleviate the feature and behavior sparsity problem. For certain factors in CTR tasks, Zheng et al. \cite{zheng_price-aware_2020} taken price into account when coming up with a GCN-based model. Furthermore, Su et al. \cite{su_detecting_2020} used L0 regularization via GNN approach to distinguish valuable factors automatically. Focusing on the re-ranking task in recommendations, Liu et al. \cite{liu_personalized_2020} developed IRGPR, which introduced an intent embedding network to embed user intents, and it is proven effective for re-ranking. Lately, several researchers have focused on applying graph neural networks to collaborative tagging systems. Chen et al. \cite{chen_tgcn_2020} used graph convolutional networks for tag-aware recommendations, and their TGCN model outperformed other state-of-the-art models. Huang et al. \cite{huang_tagnn_2021} used two graph attention networks for embeddings aggregation and achieved high-quality recommendations as well.

Few researchers use graph neural networks on tag-aware recommendations, while the model structure is quite complicated, making the training rather tricky. Our proposed method uses a relatively light and straightforward graph structure which suppose lowers the training cost and improves the performance. 

\section{Material}
\label{sec:material}
\subsection{Problem Formulation}
Folksonomy also named user tagging behavior, is the fundamental factor of the TRS. It is defined as a series of tags assigned by some users when interacting with certain items they are interested in. Generally, users interacting with certain items by operations such as clicking, tagging, or commenting could be viewed as folksonomy records. It is aggregated into a triplet, i.e, $a=(u,t,i)$, which represents user $u$ assigned tag $t$ to item $i$. These personalized tags reflect users' subjective preferences and characteristics of items. This tagging procedure is rich in collaborative information. By exploring this collaborative information, TRS can further infer users' preferences, summarize features of items and understand the connotation of tags, which hopefully improve the quality of recommendation systems.

Suppose the number of elements in user set $\mathcal{U}$, item set $\mathcal{I}$, tag set $\mathcal{T}$ are $N_u$, $N_i$ and $N_t$, respectively. The folksonomy is a tuple $\mathcal{F=(\mathcal{U},\mathcal{T},\mathcal{I},\mathcal{A})}$, where $\mathcal{A} \subseteq \mathcal{U} \times \mathcal{I} \times \mathcal{T}$ is a record in a typical TRS. During the tagging procedure, user $u$ interacts with the target item $i$ through an explicit tagging feedback by tag $t$, i.e., the user watching the movie 'Transformers' because of the tag 'sci-fi'.But few researchers have addressed the problem of information leaks on folksonomy. We argue that directly modeling explicit tagging feedback as implicit feedback may leak information because a part of tagging behavior is negative feedback, i,e, user tagging movie as 'boring movies'. The leak is supposed to hinder the recommendation performance in the test dataset. Our approach to tackling the problem is to model user-tag tagging interactions and item-tag tagged interactions in $\mathcal{A}$ separately, rather than model user-item interactions directly.

This paper focuses on recommending the personalized ranking list of items for each user $u$ on the TRS. By exploring the implicit feedback in user tagging assignments, leveraging collaborative information, and training a model to refine embeddings, we can generate the Top-K list of items for each user.

\begin{equation}
    Top(u, K) = \mathop{argmax}^{(K)}_{i\in \mathcal{I}}\hat{y}_{u,i}
\end{equation}

\subsection{Folksonomies Graph}
To prevent the information leak from happening, we construct two sets of edges $\mathcal{E}_{u,t}$ and $\mathcal{E}_{i,t}$ among the folksonomies records $a=(u,t,i) \in \mathcal{A}$. Set $\mathcal{E}_{u,t}$ reflects the assignments between users and tags. On the one hand, set $\mathcal{E}_{i,t}$ indicates the passive tagged interactions between items and tags. In TRS scenario, each kind of edge $e$ from $\mathcal{E}_{u,t}$ and $\mathcal{E}_{i,t}$ is respectively defined as: 

\begin{equation}
    e_{u,t} = 
    \begin{cases}
        1,&\text{if $(u,t) \in \mathcal{E}_{u,t}$} \\
        0,&\text{otherwise}.
    \end{cases} 
\end{equation}
\begin{equation}
    e_{i,t} = 
    \begin{cases}
        1,&\text{if $(i,t) \in \mathcal{E}_{i,t}$} \\
        0,&\text{otherwise}.
    \end{cases}
\end{equation}

\begin{figure}[H]
    \centering
    \includegraphics[width=0.9\textwidth]{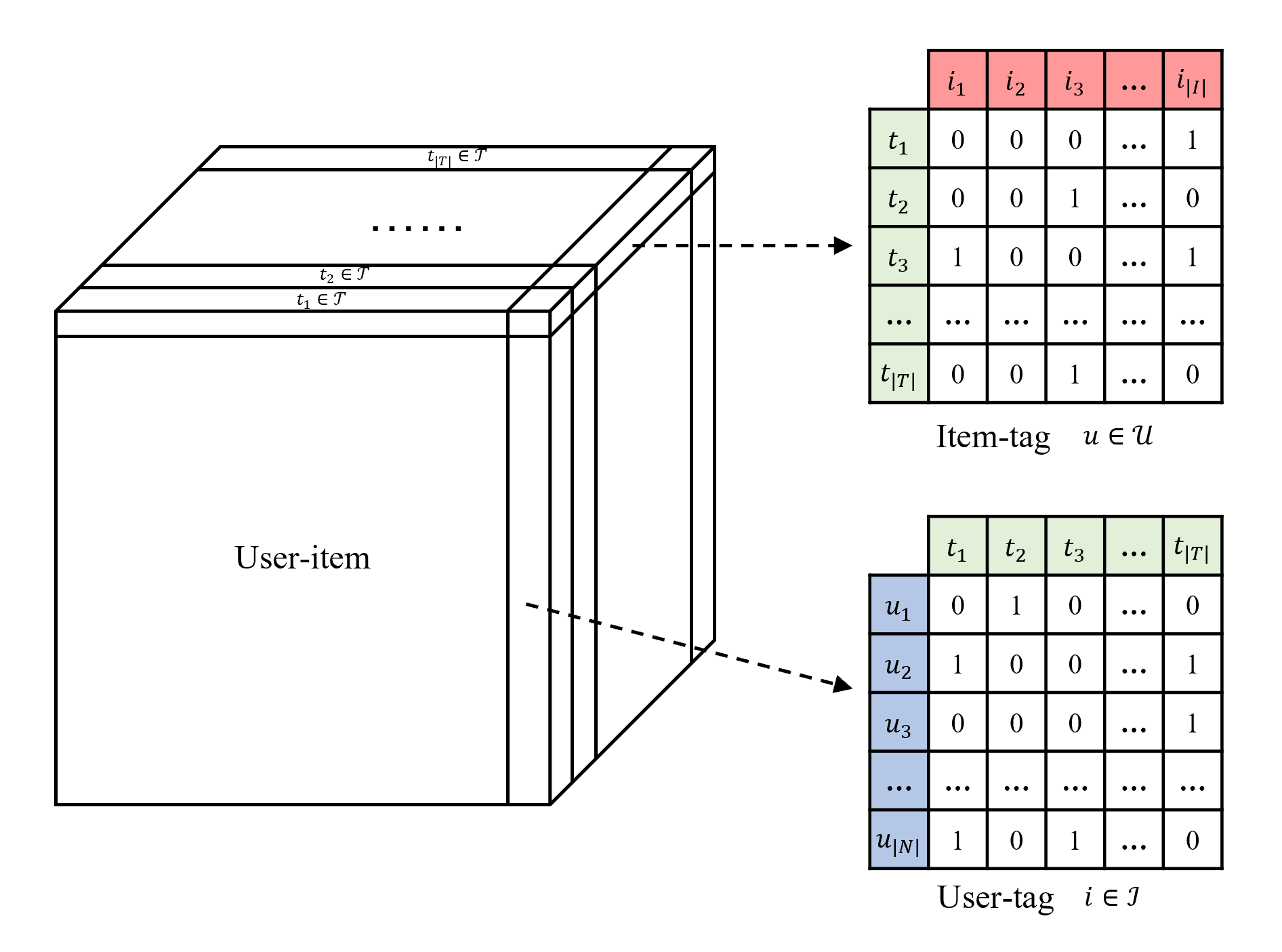}
    \caption{Matrix form of Folksonomy}
    \label{folksonomy}
\end{figure}

To make it easier to follow, the matrix form of folksonomy is shown in Fig. \ref{folksonomy}. Therefore, the FG is constructed based on user-tag tagging and item-tag tagged interactions according to the assignments set $\mathcal{A}$. Each set of edges can be respectively regarded as the set of edges in the bipartite graph $G_{tagging} = (\mathcal{U}, \mathcal{T}, \mathcal{E}_{u,t})$, $G_{tagged} = (\mathcal{I}, \mathcal{T}, \mathcal{E}_{i,t})$. 

\section{Method}
\label{sec:method}
\begin{figure}[H]
    \centering
    \includegraphics[width=\textwidth]{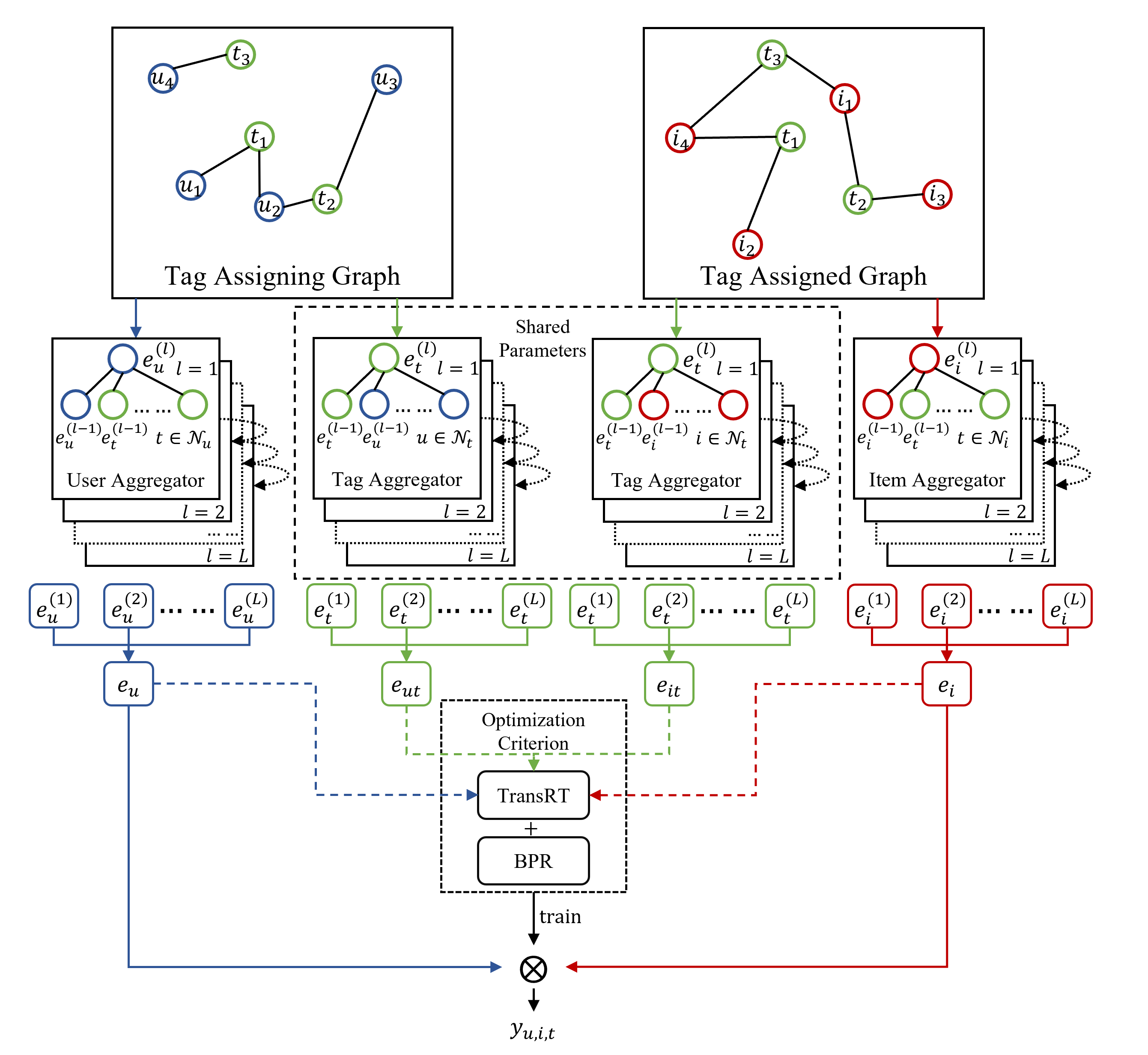}
    \caption{Model structure of LFGCF}
    \label{lagcf}
\end{figure}

In this section, we first present the design of the Light Folksonomy Graph Collaborative Filtering (LFGCF) method, as illustrated in Fig.\ref{lagcf}, which is composed of two core modules: 1) \textbf{Light Folksonomy Graph Convolutional Network}, which leverages a light yet effective model by including the essential ingredients of GCN for constructing the FG from folksonomies records $\mathcal{A}$ and GCN-based collaborative aggregated operations to capture higher-order semantic information under tagging graph and tagged graph, respectively; 2) \textbf{TransR on Tags}, which provides a regularization function named TransRT to bridge tagging graph and tagged graph by triplets preserved. The joint optimization details and how to do modeling training for Top-K recommendation in TRS will be discussed later in this section.

 \subsection{LFGCF}
Mainstream GNN models, such as GCN\cite{kipf_gcn_2017} and GAT\cite{peter_gan_2017} were originally proposed for node or graph representation learning on attributed graphs. Specifically, each node has existing embeddings as input features, which are firstly transformed into a uniform shape by features transformation and then aggregated with its neighbors on the graph. In the end, embeddings are updated by nonlinear activations. Whereas from the view of the bipartite graph for Collaborative Filtering, each node (user or item) is only identified by a unique token without any concrete semantics as input features. In such a case, given unique token embeddings as input, performing multiple layers of feature transformation and nonlinear activations are keys to the success of modern neural networks\cite{he_deep_2015}. However, these operations not only make little benefits to the performance of recommendation tasks but also increase the difficulties of representation training. Inspired by the ideas of LightGCN\cite{he_lightgcn_2020}, we propose a light yet effective model by including the essential ingredients of GCN for a recommendation.

\subsubsection{Light Assign Graph Convolutional Network}
The fundamental factor of GCN is learning nodes recommendation by aggregating neighbors features over the graph\cite{kipf_gcn_2017}. To be specific, aggregating the features of neighbors is the new representation of nodes. The neighbors aggregated function can be universally defined as follows:

\begin{equation}
    e_{u/i}^{(k+1)} = \mathcal{AGG}(e_{u/i}^{(k)}, {e_t^{(k)}: t \in \mathcal{N}_{u/i}}) 
\end{equation}
where the $\mathcal{AGG}$ is the neighbors aggregation function, which takes $k$-th layer's representations of target nodes and their neighbors into consideration. Besides, the $e_{u/i} \in \mathbf{R}^d$ is the representations of users or item embeddings in $G_{tagging}$ and $G_{tagged}$, respectively; the $e_t \in \mathbf{R}^d$ is the representations of tags and shared parameters between $G_{tagging}$ and $G_{tagged}$. Many researchers have proposed different aggregation functions for neighbors aggregation, such as the weighted sum aggregator in GIN\cite{xu_gin_2018}, LSTM aggregator in GraphSAGE\cite{hamilton_graphsage_2017}, and bilinear interaction in BGNN\cite{zhu_bgnn_2020}. However, most of them tie feature transformation or nonlinear activation with $\mathcal{AGG}$ function, such as complex attention-based and CNN-based feature transformation in TGCN\cite{chen_tgcn_2020}. Although they all perform well on node or graph classification tasks with semantics input features and TRS recommendation tasks that only have token embeddings as input features, they could be burdensome for Top-K recommendation tasks.

\subsubsection{Light Aggregation}
We leverage the LFGCN in FG, which is constructed based on $G_{tagging}$ and $G_{tagged}$. In LFCGN, we focus on the essential ingredients of GCN for recommendations. We adopt the light-weighted sum aggregator and abandon the use of feature transformation and nonlinear activation. The Light Aggregation function can be defined as follows:

\begin{eqnarray}
    e_{u/i}^{(k+1)} &=& \sum_{t\in \mathcal{N}_{u/i}}\frac{1}{\sqrt{|\mathcal{N}_{u/i}|}\sqrt{|\mathcal{N}_{t}|}}e_t^{(k)} 
    \\
    e_t^{(k+1)} &=& \sum_{u\in \mathcal{N}_t}\frac{1}{\sqrt{|\mathcal{N}_t|}\sqrt{|\mathcal{N}_{u/i}|}}e_{u/i}^{(k)}
\end{eqnarray}
The symmetric normalization term $\frac{1}{\sqrt{|\mathcal{N}_{u/i}|}\sqrt{|\mathcal{N}_t|}}$ follows the design of standard GCN\cite{kipf_gcn_2017}, which can avoid the scale of embeddings increasing with graph convolution operations.

\subsubsection{Layer Combination and Model Prediction}
In our LFGCF, the only trainable parameters are the embeddings at the 0-layer. When thay are initialize, the higher layers can be computed via LFGCN defined by Equation (5-6). After K layers LFGCN, we propose layer combination function that further combine the embeddings obtained at each layer to form the final representation of nodes. The layer combination function can be defined as follows:

\begin{equation}
    e_{u/i/t} =\mathcal{COMB}(e_{u/i/t}^{(k)}:k \in K)
\end{equation}
which $\mathcal{COMB}$ is the layer combination function, which fusion all layers' representation of specific type of nodes. $K$ is the number of layers. 

The embeddings at different layers capture different semantics in FG. For example, the first layer 
enforces smoothness on users(items) and tags that have interactions, the second layer smooths users(items) that have overlap on interacted tags, and higher-layers capture higher-order proximity\cite{wang_neural_2019}. Thus, we do not design spectial componet, and the layer combination function can be further defined as follows:
\begin{equation}
    e_u= \sum_{k=0}^{K}a_ke_u^{(k)},\quad e_i = \sum_{k=0}^{K}a_ke_i^{(k)}, \quad e_t = \sum_{k=0}^{K}a_ke_t^{(k)} 
\end{equation}
which $a_k \geq 0$ denotes the importance of the k-th layer embedding in constituting the final embedding. It can be treated as a hyperparameter to be tuned manually, or a model parameter optimized automatically. We set $a_k$ uniformly as $1/(K + 1)$, where $K$ is the number of layers. The reasons that we designded the layer combination function to get final representations are two-fold. (1) With the increasing of layers, the embeddings will be over-smoothed\cite{li_deeper_2018}. Thus only using the last layer is problematic. (2) Combining embeddings at different layers with different weight captures the effect of GCN with self-connections\cite{he_lightgcn_2020}. 

The model prediction is defined as the inner product of the user and item final representations:

\begin{equation}
    \hat{y}_{ui} = e_u^{T}e_i
\end{equation}
where is used as the ranking score for recommendation generation.

 \subsection{TransRT}
Knowledge graph embedding effectively parameterizes nodes as vector representations while keeping the graph's topology. In this paper, we propose a new method for embedding a knowledge graph based on the idea of transformers. Here we propose a new regularization function, which is based on TransR\cite{lin_learning_nodate}, a widely used method in a knowledge graph. To be specific, it learns the embedding of each node by optimizing the translation principle $e_u + e_t \approx e_t$, if a triplet $(u,t,i) \in \mathcal{A}$. Herein, $e_u, e_i, e_t \in \mathbf{R}^d$ is the final embedding of user $u$, item $i$ and tag $t$, respectively, and $e_u, e_i$ are the projected representations of $e_u$ and $e_i$ in the tag's space. Hence, for a given folksonomy record $(u, t, i)$, its plausibility score could be defined as follows:

\begin{equation}
    g(u, t, i) = ||e_u + e_t - e_i||_2^2
\end{equation}
where $e_u, e_i, e_t$ are in the same d-dimension space, but not the same semantics space. A lower score of $g(u, t, i)$ suggests that the folksonomy score is more likely to be reasonable and vice versa.
 \subsection{Jointly Optimization Details}
The trainable parameters of LFGCF are only the embeddings of the 0-th layer, which combine with users, items, and tags in FG. In other words, the model complexity is the same as the standard matrix factorization (MF). To optain better ranking, we employ the Bayesian Personalized Ranking (BPR) loss\cite{rendle_bpr_2009}, which is a pairwise loss that encourages the prediction of an observed record to be higher than its unobserved sampled counterpart. The BPR loss is defined as follows:

\begin{equation}
    \mathcal{L}_{LFGCN} = \sum_{(u,i,i') \in \mathcal{O}} -ln(\sigma(\hat{y}_{u,i} - \hat{y}_{u,i'}))
\end{equation}
where $\mathcal{O} = \{(u,i,i')|(u,i) \in \mathcal{A},(u,i') \notin \mathcal{A}\}$ indicates pairwsie $(u,i)$ observed in folksonomy records, and pairwise $(u,i')$ means that the user $u$ and item $i'$ not observed in record, but uniformly sampled from the unobserved pairs.

To train TransRT, we minimize the plausibility score:

\begin{equation}
    \mathcal{L}_{TranRT} = \alpha g(u, t, i)
\end{equation}
where $\alpha$ controls the strength of the knowledge graph regularization, and $g(u, t, i)$ is the plausibility score of the record $(u, t, i)$.

To effective learning parameters for recommendation and preserve the regularization relationship among folksonomy records, we integrate the Top-K recommendation task and the TransRT by a jointly learning framework. Finally, the total objective function of LFGCF is defined as follows:

\begin{equation}
    \mathcal{L}_{LFGCF} = \mathcal{L}_{LFGCN} + \mathcal{L}_{TranRT} + \gamma ||\Theta||_2
\end{equation}
where $\gamma$ controls the strength of regularization. We employ the Adam\cite{kingma_adam_2014} optimize $\mathcal{L}_{LFGCN}$ and $\mathcal{L}_{TranRT}$ and use it in a mini-batch manner. Besides, an early stopping strategy is also applied to avoid overfitting during training.
  
\section{Experimental}
\label{sec:experiment}
In this section, we focus on the following questions:

\subparagraph{RQ1} Does LFGCF outperform other tag-aware recommendation models in the Top-K recommendation task?

\subparagraph{RQ2} Does it help improve the recommendation performance to remove some components in the GCN?

\subparagraph{RQ3} Whether or not the implementation of TransRT solves the redundancy and ambiguity of tags?

\subsection{Setup}

In order to answer the questions above, a series of experiments are designed and carried out. We first show that LFGCF outperforms other state-of-the-art models, then elaborate on its expressiveness and explainability. Meanwhile, according to the investigation from \cite{dacrema_arewe_2019}, experiments of some other models are carried out with the different processes, which brings difficulties to repeating them. Hence, all the experiments in this paper are all under the uniformed experimental framework Recbole\cite{zhao_recbole_2021} to make fair comparisons.

\subsubsection{Experiment Datasets}

Extensive experiments are carried out to evaluate the proposed LFGCF based on three real-world datasets: MovieLens, LastFM, and Delicious. They were all released in HetRec \cite{cantador_hetrec_2011}. 

\begin{itemize}
    \item [$\bullet$] \textbf{MovieLens} is a recommendation dataset that contains a list of tags assigned by users to various movies.
    \item [$\bullet$] \textbf{LastFM} is a dataset from Last.FM music system with music listening information, tags assignments to artists, and user social networks.
    \item [$\bullet$] \textbf{Delicious} is obtained from Del.icio.us and contains tagging information to web bookmarks.
\end{itemize}

Due to the sparsity of tagging information, some infrequently used tags exist. To rule out their negative influence of them, those tags used less than 5 times in MovieLens and LastFM and 15 times in Delicious are removed\cite{zuo_tag-aware_2016}. Basic statistic information of the  datasets after preprocessing is summarized in Table.\ref{data_sta}

\begin{table}[h]
    \centering
    \caption{Datasets Statistics}
      \begin{tabular}{c|c|c|c|c|c}
      \hline
      \textbf{Datasets} & \textbf{User} & \textbf{Item} & \textbf{Tag} & \textbf{Assignments} & \textbf{Sparsity} \\
      \hline
      Last.FM & 1808 & 12212 & 2305 & 175641 & 99.20\% \\
      MovieLens & 1651 & 5381 & 1586 & 36728 & 99.59\% \\
      Delicious & 1843 & 65877 & 3508 & 330744 & 99.73\% \\
      \hline
      \end{tabular}
    \label{data_sta}
  \end{table}

Since data we use are not time sequence data, so training sets, validation sets and test sets are randomly selected according to the proportion of \{0.6, 0.2, 0.2\}. The metrics that reflect model performances in the reminder part of the paper are all calculated from test sets.

\subsection{Evaluation Metrics}

The performance of TRS is directly related to the quality of Top-K recommendations, which are evaluated by the following metrics: Recall@N, Precision@N, Hit Ratio@N, NDCG@N, and MRR@N\cite{jarvelin_cumulated_2002}. Empirically, higher metrics mean better performances. Each metric is  elaborated:
\begin{itemize}
  \item [$\bullet$] \textbf{Recall@N} measures the percentage of the number of items in Top-K recommendations to the actual item set that user interact with. 
  
  \begin{equation}
    Recall@N = \frac{|R^N(u) \bigcap T(u)|}{|T(u)|}
  \end{equation}
  where $R^N(u)$ denotes the Top-K recommendations and $T(u)$ denotes the ground truth item set.

  \item [$\bullet$] \textbf{Precision@N} measures the fraction of items the users would interact with among the Top-K recommendations.
  
  \begin{equation}
    Precision@N = \frac{|R^N(u) \bigcap T(u)|}{K}
  \end{equation}

  \item [$\bullet$] \textbf{Hit Ratio@N} measures the percentage of users who interact with the recommendations for at least one time.
  
  \begin{equation}
    HR@N = \frac{1}{\mathcal{U}}\sum_{u \in \mathcal{U}}I(|R^N(u) \bigcap T(u)|>0)
  \end{equation}
  where $I(\cdot)$ is the indicator function.
  \item [$\bullet$] \textbf{NDCG@N} reflects the quality of ranking by distinguishing the contributions of the accurately recommended items.
  
  \begin{equation}
    NDCG@N = \frac{1}{\mathcal{U}}\sum_{u \in \mathcal{U}} \frac{\sum^N_{n=1} \frac{I(R^N_n(u) \in T(u))}{log(n+1)}}{\sum^N_{n=1}\frac{1}{log(n+1)}}
  \end{equation}
  where $R^N_n(u)$ means the $n^{th}$ item in Top-K recommendations $R^N(u)$.
  \item [$\bullet$] \textbf{MRR@N} computes the reciprocal rank of the first relevant item found by an rank algorithm.
  
  \begin{equation}
    MRR@N = \frac{1}{\mathcal{U}} \sum_{u \in \mathcal{U}}\frac{1}{rank_u^*}
  \end{equation}
  where $rank_u^*$ means the rank position of the first relevant item in recommendations for a user. 
\end{itemize}

\subsection{Baselines and Parameters}

To fairly evaluate the performance and effectiveness of LFGCF, we adopt some classic or state-of-the-art TRS models as baselines.

\begin{itemize}
    \item [$\bullet$] \textbf{DSPR}\cite{xu_dspr_2016} leverages deep neural networks to learn tag-based features by mapping user and item profiles to deep latent space.
    \item [$\bullet$] \textbf{CFA}\cite{zuo_tag-aware_2016} is a user-based collaborative filtering model which adopts a sparse autoencoder to extract latent features.
    \item [$\bullet$] \textbf{BPR-T}\cite{li_tag-aware_2019} is a collaborative filtering model which incorporates tagging information and the Bayesian ranking optimization.
    \item [$\bullet$] \textbf{TGCN}\cite{chen_tgcn_2020} is a collaborative filtering model which incorporates tagging information into GCN along with an attention mechanism.
\end{itemize}

In order to make impartial comparisons, each model is optimized with mini-batch Adam, while the batch size is set as 2048. The learning rate of each model is searched from \{0.0001, 0.0005, 0.001, 0.005, 0.01, 0.05\} and the regression weight is searched from \{1e-5, 1e-4, 1e-3, 1e-2\}. For the autoencoder in CFA and the graph structure in TGCN and LFGCF, the number of layers is searched from \{1, 2, 3, 4\}. For BPR-T, its three regression weight is searched from \{1e-5, 1e-4, 1e-3, 1e-2, 1e-1\}. Additionally, by further searching the coefficients $\alpha, \beta$ from \{1e-5, 1e-4\}, we can analyze the sensibility of TransRT. The hyperparameter experiments are conducted under 10 random seeds. By conducting these experiments, all of the models are at their optimal performances, ensuring the fairness of the following comparisons. %
\subsection{Performance Analysis}

The experiment results in this section are all achieved with the optimal hyperparameters. Best performance is in boldface, best baseline proformance is in underline and imp. means the improvement of LFGCF over the state-of-the-art baseline.
\newpage

\begin{center}
    \begin{longtable}{c|c|c|c|c|c|c|c}
      \caption{Performance Comparison}
      \label{performance_comparison}\\
        \hline
        \textbf{Dataset} & \textbf{Metric} & \textbf{DSPR} & \textbf{CFA} & \textbf{BPRT} & \textbf{TGCN} & \textbf{LFGCF} & \textbf{imp}\\
        \hline
          MovieLens & \begin{tabular}[c]{@{}c@{}}
            Recall@10 \\ Recall@20 \\ Precision@10 \\ Precision@20 \\ Hit@10 \\ Hit@20 \\ NDCG@10 \\ NDCG@20 \\ MRR@10 \\MRR@20
          \end{tabular} 
          & \begin{tabular}[c]{@{}l@{}}
            0.0755 \\ 0.1076 \\ 0.0326 \\ 0.0266 \\ 0.1838 \\ 0.2358 \\ 0.0683 \\ 0.0748 \\ 0.0949 \\ 0.0984
          \end{tabular} 
          & \begin{tabular}[c]{@{}l@{}}
            0.0549 \\ 0.1073 \\ 0.0180 \\ 0.0169 \\ 0.1269 \\ 0.2068 \\ 0.0397 \\ 0.0533 \\ 0.0549 \\ 0.0607
          \end{tabular} 
          & \begin{tabular}[c]{@{}l@{}}
            \underline{0.2032} \\ 0.2146 \\ 0.0353 \\ 0.0256 \\ 0.2773 \\ 0.3025 \\ \underline{0.1778} \\ \underline{0.1822} \\ \underline{0.1879} \\ \underline{0.1894}
          \end{tabular} 
          & \begin{tabular}[c]{@{}l@{}}
            0.1951 \\ \underline{0.2362} \\ \underline{0.0400} \\ \underline{0.0292} \\ \underline{0.2997} \\ \underline{0.3626} \\ 0.1639 \\ 0.1748 \\ 0.1790 \\ 0.1833
          \end{tabular} 
          & \begin{tabular}[c]{@{}l@{}}
            \textbf{0.2928} \\ \textbf{0.3035} \\ \textbf{0.0546} \\ \textbf{0.0345} \\ \textbf{0.3950} \\ \textbf{0.4118} \\ \textbf{0.2468} \\ \textbf{0.2489} \\ \textbf{0.2482} \\ \textbf{0.2492}
          \end{tabular}
          & \begin{tabular}[c]{@{}l@{}}
            44.09\% \\ 28.49\% \\ 36.5\% \\ 18.15\% \\ 31.80\% \\ 13.57\% \\ 38.81\% \\ 36.61\% \\ 32.09\% \\ 31.57\%
          \end{tabular} \\
        \hline
        Last.FM & \begin{tabular}[c]{@{}c@{}}
            Recall@10 \\ Recall@20 \\ Precision @10 \\ Precision@20 \\ Hit@10 \\ Hit@20 \\ NDCG@10 \\ NDCG@20 \\ MRR@10 \\MRR@20
          \end{tabular} 
          & \begin{tabular}[c]{@{}l@{}}
            0.0982 \\ 0.1472 \\ 0.0971 \\ 0.0754 \\ 0.3883 \\ 0.4637 \\ 0.1373 \\ 0.1389 \\ 0.2204 \\ 0.2257
          \end{tabular} 
          & \begin{tabular}[c]{@{}l@{}}
            0.1086 \\ 0.1661 \\ 0.0727 \\ 0.0603 \\ 0.3596 \\ 0.4561 \\ 0.1220 \\ 0.1320 \\ 0.1876 \\ 0.1946
          \end{tabular} 
          & \begin{tabular}[c]{@{}l@{}}
            0.3357 \\ 0.4118 \\ 0.1371 \\ 0.1106 \\ 0.6649 \\ 0.7450 \\ 0.3317 \\ 0.3487 \\ 0.4080 \\ 0.4136
          \end{tabular} 
          & \begin{tabular}[c]{@{}l@{}}
            \underline{0.3745} \\ \underline{0.4444}\\ \underline{0.1691} \\ \underline{0.1325} \\ \underline{0.6971} \\ \underline{0.7649} \\ \underline{0.3920} \\ \underline{0.4044} \\ \underline{0.4708} \\ \underline{0.4755}
          \end{tabular} 
          & \begin{tabular}[c]{@{}l@{}}
            \textbf{0.4189} \\ \textbf{0.4980} \\ \textbf{0.1847} \\ \textbf{0.1475} \\ \textbf{0.7404} \\ \textbf{0.7889} \\ \textbf{0.4248} \\ \textbf{0.4442} \\ \textbf{0.5023} \\ \textbf{0.5058}
          \end{tabular}
          & \begin{tabular}[c]{@{}l@{}}
            11.86\% \\ 12.06\% \\ 9.22\% \\ 11.32\% \\ 6.21\% \\ 3.13\% \\ 8.36\% \\ 9.84\% \\ 6.69\% \\ 6.37\%
          \end{tabular} \\
        \hline
        Delicious & \begin{tabular}[c]{@{}c@{}}
            Recall@10 \\ Recall@20 \\ Precision @10 \\ Precision@20 \\ Hit@10 \\ Hit@20 \\ NDCG@10 \\ NDCG@20 \\ MRR@10 \\MRR@20
          \end{tabular} 
          & \begin{tabular}[c]{@{}l@{}}
            0.0134 \\ 0.0195 \\ 0.0320 \\ 0.0262 \\ 0.1611 \\ 0.2284 \\ 0.0374 \\ 0.0335 \\ 0.0788 \\ 0.0836
          \end{tabular} 
          & \begin{tabular}[c]{@{}l@{}}
            0.0097 \\ 0.0182 \\ 0.0112 \\ 0.0102 \\ 0.0963 \\ 0.1526 \\ 0.0161 \\ 0.0173 \\ 0.0377 \\ 0.0415
          \end{tabular} 
          & \begin{tabular}[c]{@{}l@{}}
            0.1396 \\ 0.2595 \\ 0.2473 \\ 0.2736 \\ 0.8069 \\ \underline{0.9174} \\ 0.2485 \\ 0.2961 \\ 0.3195 \\ 0.3277
          \end{tabular} 
          & \begin{tabular}[c]{@{}l@{}}
            \underline{0.1769} \\ \underline{0.3073} \\ \underline{0.3586} \\ \underline{0.3514} \\ \underline{0.8835} \\ 0.9168 \\ \underline{0.3866} \\ \underline{0.4048} \\ \underline{0.5474} \\ \underline{0.5498}
          \end{tabular} 
          & \begin{tabular}[c]{@{}l@{}}
            \textbf{0.1956} \\ \textbf{0.3341} \\ \textbf{0.3716} \\ \textbf{0.3615} \\ \textbf{0.9032} \\ \textbf{0.9404} \\ \textbf{0.4015} \\ \textbf{0.4225} \\ \textbf{0.5546} \\ \textbf{0.5572}
          \end{tabular}
          & \begin{tabular}[c]{@{}l@{}}
            10.57\% \\ 8.72\% \\ 3.63\% \\ 2.87\% \\ 2.23\% \\ 2.51\% \\ 3.85\% \\ 4.37\% \\ 1.32\% \\ 1.35\%
          \end{tabular} \\
        \hline
    \end{longtable}
    \end{center}

\begin{figure}[!htb]
    \centering
    \includegraphics[width=\textwidth]{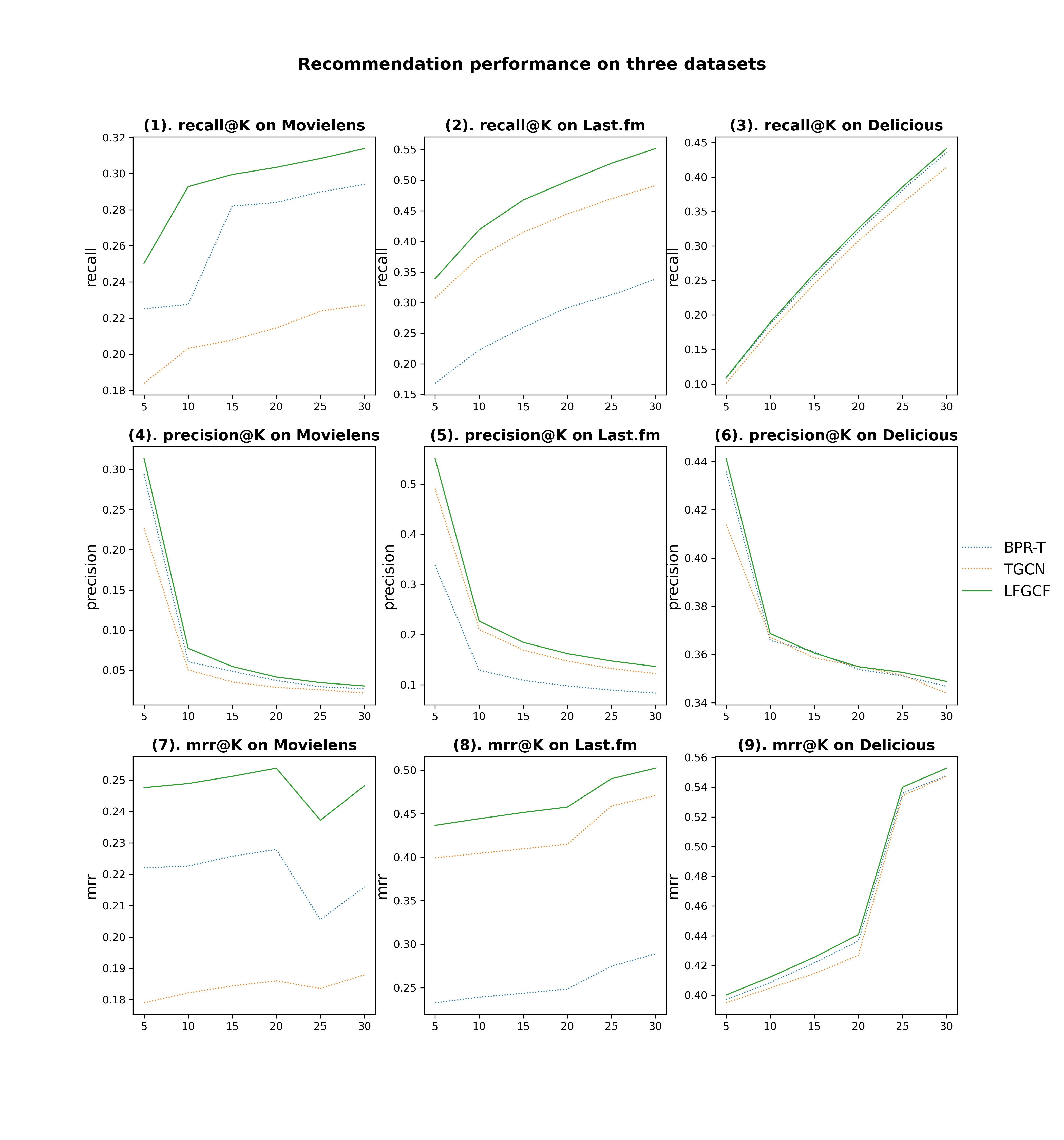}
    \caption{Performace comparison on three datasets}
    \label{per_com}
\end{figure}

Table.\ref{performance_comparison} shows the Top-K recommendation performance metrics of LFGCF and other baselines in three datasets when $N = \{10, 20\}$. Fig.\ref{per_com} shows the Top-K recommendation performance of our LFGCF and other baselines in terms of Recall@N, Precision@N and MRR@N while N ranges from 5 to 30 with the interval of 5.

Fig.\ref{per_com} shows the Top-K recommendation performance of our LFGCF and other baselines in Recall@N, Precision@N, and MRR@N, while N ranges from 5 to 30 with an interval of 5.

The performance comparisons of our model and other baselines shows that LFGCF achieved the state-of-the-art performance while succcessfully alleviating the training difficulty with the help of light designed GCN. Among baseline models, TGCN and BPR-T outperform DSPR and CFA by a large margin. 

We summarize why our LFGCF outperforms other baselines for the following reasons: (1) More effective GCN. Two LFGCN implemented for feature learning show better performances than deep neural networks in models CFA and DSPR. Effective representation propagation and aggregation benefit LFGCF by lifting performances and reducing training difficulties; (2) TransRT assists with extracting more expressive user and item representations. Recent years witnessed efforts devoted to the loss function based on BPR loss to better fit recommendation tasks. Experiments indicate that using the improved knowledge graph algorithm TransRT allows more efficient uses of collaborative tagging information. %
\subsection{Ablation Studies}

To verify our summaries of the superior performance of LFGCF above, we conduct a series of ablation studies on LFGCF.
\subsubsection{Effect of LFGCN}

Some models take user-item interaction information into personalized recommendations in the research field. According to the research of He et al. \cite{he_lightgcn_2020}, complex feature transformation and nonlinear activation not only bring difficulties to training but also compromise the performance. To verify the effectiveness of light designed GCN in tag-aware recommendation systems, We implement NGCFT as a baseline model based on NGCF \cite{wang_neural_2019}. Apart from the complex feature aggregation and propagation operations inherited from NGCF, NGCFT is identical to LFGCF in the loss function, recommendation generation, and parameter setting.

\begin{center}
    \begin{longtable}{c|c|c|c|c|c}
        \caption{Performance of NGCFT and LFGCF on Movielens}
        \label{ngcft_ml}\\
          \hline
          \textbf{Model} & \textbf{Recall@20} & \textbf{Precision@20} & \textbf{Hit@20} & \textbf{NDCG@20} & \textbf{MRR@20} \\
          \hline
          NGCFT & 0.2158 & 0.0227 & 0.3109 & 0.1383 & 0.1405 \\
          LFGCF & 0.2767 & 0.0375 & 0.4176 & 0.2156 & 0.2199 \\
          imp & 28.22\% & 65.20\% & 34.32\% & 55.89\% & 56.51\% \\
          \hline
    \end{longtable}
\end{center}

\begin{center}
    \begin{longtable}{c|c|c|c|c|c}
        \caption{Performance of NGCFT and LFGCF on LastFM}
        \label{ngcft_lfm}\\
          \hline
          \textbf{Model} & \textbf{Recall@20} & \textbf{Precision@20} & \textbf{Hit@20} & \textbf{NDCG@20} & \textbf{MRR@20} \\
          \hline
          NGCFT & 0.5005 & 0.1454 & 0.7953 & 0.4543 & 0.5167 \\
          LFGCF & 0.5069 & 0.1468 & 0.8035 & 0.4442 & 0.5058 \\
          imp & 1.28\% & 0.96\% & 1.03\% & -2.23\% & -2.11\% \\
          \hline
    \end{longtable}
\end{center}

\begin{center}
    \begin{longtable}{c|c|c|c|c|c}
        \caption{Performance of NGCFT and LFGCF on Movielens}
        \label{ngcft_de}\\
          \hline
          \textbf{Model} & \textbf{Recall@20} & \textbf{Precision@20} & \textbf{Hit@20} & \textbf{NDCG@20} & \textbf{MRR@20} \\
          \hline
          NGCFT & 0.3256 & 0.3574 & 0.9354 & 0.4150 & 0.5453 \\
          LFGCF & 0.3341 & 0.3615 & 0.9404 & 0.4225 & 0.5572 \\
          imp & 2.61\% & 1.15\% & 0.53\% & 1.81\% & 2.18\% \\
          \hline
    \end{longtable}
\end{center}

Generally, LFGCF outperforms NGCFT on three datasets. On MovieLens, LFGCF outperforms NGCFT by a large margin. This suggests that using the light graph structure effectively imporves the performance of recommendations.

\subsubsection{Effect of TransRT}

Modified knowledge graph algorithm TransRT is implemented along with BPR loss to help with model training and achieving more expressive representations. Theoretically speaking, TransRT allows for better-using tagging information to improve qualities of user and item representations, which would lift the performance of recommendations. To verify the effectiveness of TransRT, we remove TransRT and only train the LFGCF with BPR loss and name this baseline model LFGCF-RT.

\begin{figure}[H]
    \centering
    \subfigure[LastFM]{
        \label{density_lastfm}
        \includegraphics[width=0.45\textwidth]{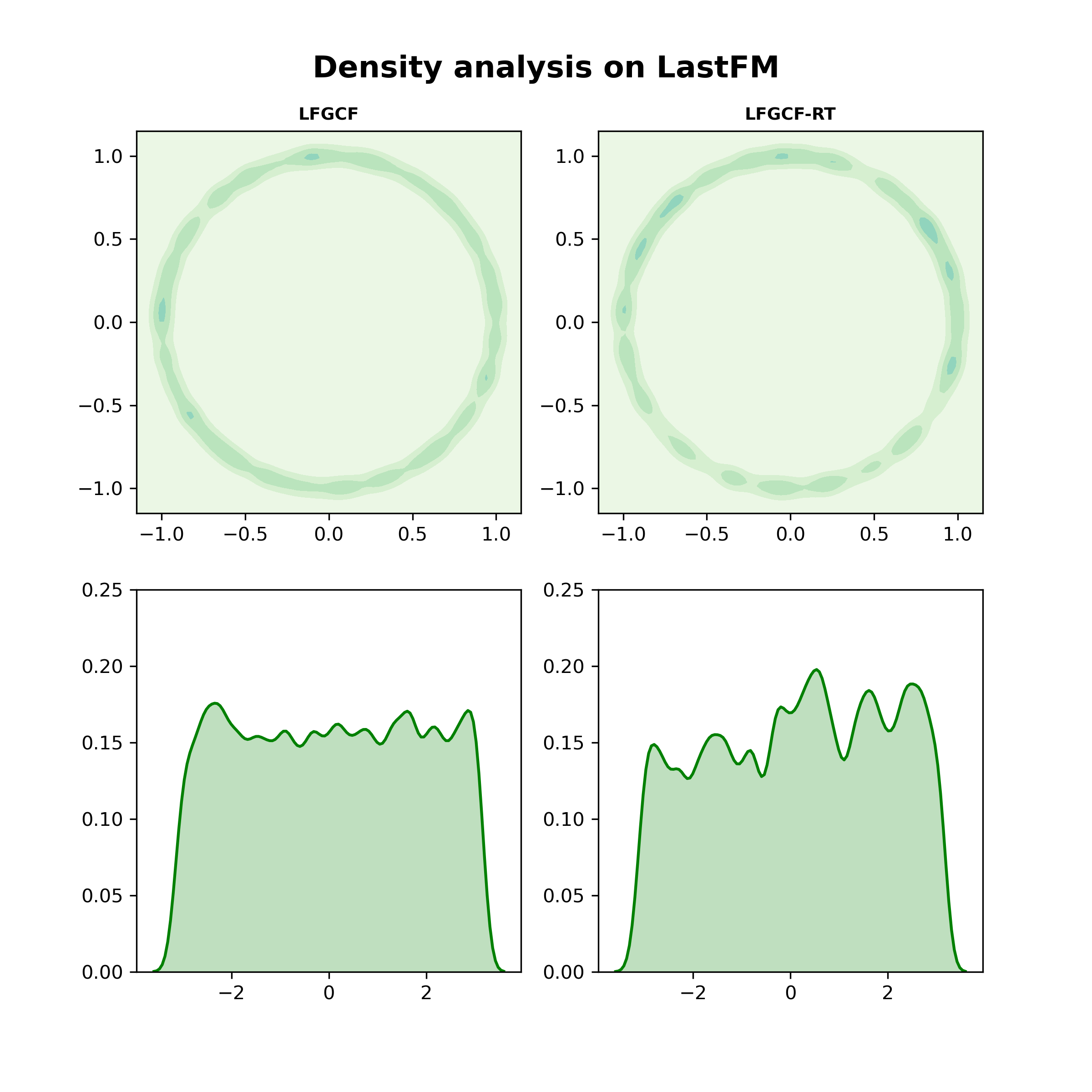}
    }
    \hspace{0in}
    \subfigure[Delicious]{
        \label{density_delicious}
        \includegraphics[width=0.45\textwidth]{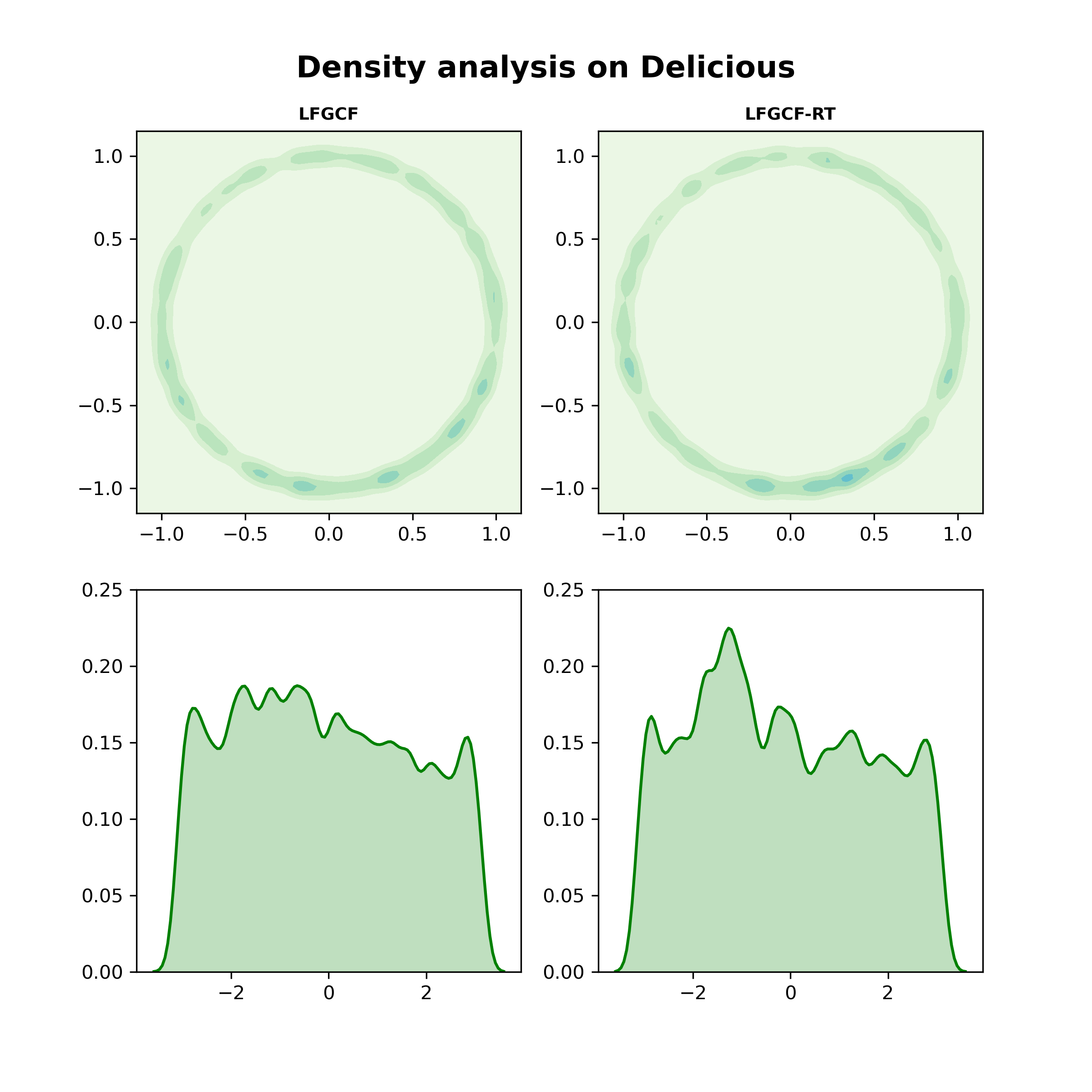}
    }
    \caption{Density analysis of LFGCF and LFGCF-RT}
    \label{density_analysis_lfm}
\end{figure}

Apart from learning more expressive user and item representations, we believe that TransRT help learn tag representations as well. To verify this deduction, we follow the visualization method in \cite{yu_are_2022}. Learned tag embeddings are first reduced to two dimensions using t-SNE \cite{van_Visualizing_2008}. Then 2-dimensional embeddings are normalized and mapped on the unit hypersphere (i.e., a circle with radius 1). To make the presentation, the density estimations on angles for each point on the hypersphere are visualized in Fig.\ref{density_analysis_lfm}.

As it can be learned from the Fig.\ref{density_analysis_lfm}, tag representations trained by LFGCF are much smoother than those learned by the baseline model without TransRT. It indicates that TransRT help promote the recommendation performance by smoothing the learned representations. To make a clear comparison between LFGCF and LFGCF-RT, the performances are shown in the Table.\ref{transrt_ml}

\newpage

\begin{center}
    \begin{longtable}{c|c|c|c|c|c}
        \caption{Performance of LFGCF-RT and LFGCF on Movielens}
        \label{transrt_ml}\\
          \hline
          \textbf{Model} & \textbf{Recall@20} & \textbf{Precision@20} & \textbf{Hit@20} & \textbf{NDCG@20} & \textbf{MRR@20} \\
          \hline
          LFGCF-RT & 0.2739 & 0.0311 & 0.3782 & 0.2109 & 0.2096 \\
          LFGCF & 0.2767 & 0.0375 & 0.4176 & 0.2156 & 0.2199 \\
          imp & 1.02\% & 20.58\% & 10.42\% & 2.23\% & 4.91\% \\
          \hline
    \end{longtable}
\end{center}

\begin{center}
    \begin{longtable}{c|c|c|c|c|c}
        \caption{Performance of LFGCF-RT and LFGCF on LastFM}
        \label{transrt_lfm}\\
          \hline
          \textbf{Model} & \textbf{Recall@20} & \textbf{Precision@20} & \textbf{Hit@20} & \textbf{NDCG@20} & \textbf{MRR@20} \\
          \hline
          LFGCF-RT & 0.4637 & 0.1358 & 0.7708 & 0.3995 & 0.4565 \\
          LFGCF & 0.5069 & 0.1468 & 0.8035 & 0.4442 & 0.5058 \\
          imp & 9.32\% & 8.10\% & 4.24\% & 11.19\% & 10.80\% \\
          \hline
    \end{longtable}
\end{center}

\begin{center}
    \begin{longtable}{c|c|c|c|c|c}
        \caption{Performance of LFGCF-RT and LFGCF on Movielens}
        \label{transrt_de}\\
          \hline
          \textbf{Model} & \textbf{Recall@20} & \textbf{Precision@20} & \textbf{Hit@20} & \textbf{NDCG@20} & \textbf{MRR@20} \\
          \hline
          LFGCF-RT & 0.3161 & 0.3512 & 0.9223 & 0.4068 & 0.5516 \\
          LFGCF & 0.3254 & 0.3526 & 0.9305 & 0.4122 & 0.5558 \\
          imp & 2.94\% & 0.40\% & 0.89\% & 1.33\% & 0.76\% \\
          \hline
    \end{longtable}
\end{center}

Table.\ref{transrt_ml} shows the recommendation performance of LFGCF and LFGCF-RT. LFGCF-RT underperforms LFGCF in all three datasets, indicating that the removal of TransRT hurts LFGCF. It compromises the performance to the same level as BPR-T and TGCN. The conclusion can be drawn that the light convolutional graph is effective in learning user and item representations, but the implicit feedbacks are not fully extracted. Implementing TransRT allows for leveraging the information pattern inside user tagging behavior to make the representations more expressive.   

\section{Conclusions and Future Work}
\label{sec:conclusions}
In this work, we explore folksonomy records with collaborative information in FG for a tag-aware recommendation. We devised a new method LFGCF, which leverages a light yet effective model to capture higher-order information in FG, then proposed a regularization function to bridge users and items in a folksonomy records triplet as its core is the GCN-based model LFGCN. It only consists of two essential components - light aggregation and information updating. In light aggregation, we remove feature transformation and nonlinear activation, which are burdensome. We construct the final representations for users and items as a weighted sum of their embeddings on several layers in information updating. For adequate modeling folksonomy records, which keep the topology information, we proposed TransRT regularization function and performed jointly learning to users' subjective preferences and items' characteristics. We argued that the specific light design and regularization for TRS alleviate the sparsity, redundancy, and ambiguity issues in folksonomy records. Extensive hyperparameter experiments and ablation studies on three real-world datasets demonstrate the effectiveness and rationality of LFGCN.

This work explores the specific design of GNN in TRS. We believe the insights of LFGCF are inspirational for future developments in TRS. With the prevalence of user tagging behaviors in real applications, GNN-based models are becoming increasingly crucial in TRS; by explicitly exploiting the interactions among entities in the system, they are advantageous to context-based supervised learning scheme\cite{he_neural_2017} that models the interactions implicitly. Besides folksonomy, much other structural information can be extracted from real-world recommendation systems, such as social networks and knowledge graphs. For example, by integrating entity relationships with FG, we can capture the explicit information rather than collaborative information. In further work, we would like to exploit further self-supervised learning for TRS, which is a promising research direction.

\end{document}